\documentclass[procediaempty]{easychair}
\usepackage{doc}
\usepackage{subfig}
\def\procediaConference{}
\title{Selection of Random Walkers that Optimizes the Global Mean First-Passage Time for Search in Complex Networks}
\titlerunning{Selection of Random Walkers that Optimizes the Searching Time in Networks}
\author{
	Mu Cong Ding\inst{1}\thanks{M.C. Ding acknowledges the support of the HKUST Undergraduate Research Opportunity Program.}
	\and
	Kwok Yip Szeto\inst{1}
}
\institute{
	Hong Kong University of Science and Technology,
	HKSAR, China\\
	\email{phszeto@ust.hk}
}
\authorrunning{Ding, Szeto}
\begin{document}
	\maketitle
	\begin{abstract}
	We design a method to optimize the global mean first-passage time (GMFPT) of multiple random walkers searching in complex networks for a general target, without specifying the property of the target node. According to the Laplace transformed formula of the GMFPT, we can equivalently minimize the overlap between the probability distribution of sites visited by the random walkers. We employ a mutation only genetic algorithm to solve this optimization problem using a population of walkers with different starting positions and a corresponding mutation matrix to modify them. The numerical experiments on two kinds of random networks (WS and BA) show satisfactory results in selecting the origins for the walkers to achieve minimum overlap. Our method thus provides guidance for setting up the search process by multiple random walkers on complex networks. 
	\end{abstract}
	%\setcounter{tocdepth}{2}
	%{\small
	%\tableofcontents}
	%Processing in EasyChair - 5 pages.
	%------------------------------------------------------------------------------
	\section{Introduction}
	\label{sect:introduction}
	Network provides a succinct mathematical representation of the interaction between components in complex systems \cite{RevModPhys.74.47}. One of the recent topics of research in network concerns the studies of information spreading. A particularly useful measure in the spreading of signal in network is the first-passage time (FPT) needed by a random walker to reach a target node starting from a chosen origin \cite{hughes1995random}. Among many papers on the mean first-passage time (MFPT) \cite{PhysRevE.75.021111}, the useful relation between the asymptotic form of the MFPT distribution and the structural properties of a network can be employed for efficient search provided that we have some prior knowledge of the target \cite{PhysRevLett.92.118701,0295-5075-90-4-40005}. However, without information about the target node, the problem of search efficiency remains an unsolved problem as we do not know how to minimize the global mean first-passage time (GMFPT) to a general target when multiple random walkers are employed. In this paper, we solve this problem by designing a method to optimize the GMFPT by choosing the initial positions of multiple random walkers.
	
	Consider there are $r$ independent random walkers on a network $G$. For each random walker taking $T$ steps, there will be a probability distribution on the set of visited sites. Intuitively, larger path distances between the initial positions of the walkers will lead to a smaller overlap between the probability distribution of the walkers. Since a smaller overlap implies there is less time wasted by the walkers, the average searching time can be shortened. This intuition is verified by the theoretical analysis on the case of two walkers. We thus like to find the set of origins of the random walkers that minimize the overlap. In this paper, we use genetic algorithm \cite{holland1975adaptation} to find an optimal set of initial nodes. We employ a special evolutionary framework with a mutation matrix, which was called the mutation only genetic algorithm (MOGA), to design the optimization algorithm. MOGA was first introduced by Szeto and Zhang \cite{Szeto2006} and later generalized by Law and Szeto to include crossover \cite{law2007adaptive,wu2014applications}. Simulations are performed on the two main kinds of random networks: WS and BA, with satisfactory results.
	%------------------------------------------------------------------------------
	\section{Global Mean First-Passage Time and Overlap}
	\label{sect:theory}
	Consider a connected and undirected finite network $G=(V,E)$ with adjacency matrix $\mathbf{A}$. The degree of a node $i$ is denoted by $k_i$. A Random walk is a discrete-time stochastic process. A walker at node $i$ will choose one of its neighbors as the next step with equal chances. For a random walker, if it starts at the initial node $i$, the chance of finding it at node $j$ at time $t$ is denoted by $P_{ij}(t)$. We may introduce the stochastic matrix $S_{ij}=A_{ij}/k_i$ and the probability distribution row vector $[\mathbf{P}_i(t)]_j=P_{ij}(t)$. Then the master equation of a random walk is $\mathbf{P}_i(t)=\mathbf{P}_i(t-1)\mathbf{S}=\mathbf{P}_i(0)\mathbf{S}^{t}$. 
	
	Our goal is to optimize the global mean first-passage time (GMFPT) while searching by multiple walkers. However, computing the GMFPT needs the global topological structure of the network and is time-consuming since it converges slowly when we simulate random walks. In order to design an algorithm to optimize the GMFPT, we first want to find a fast computable quantity that is monotonically related to it. Following our intuition that an overlap between the probability distributions of distinct walkers is a waste of searching time, we expect that a smaller overlap may result in a smaller GMFPT. 
	
    Now we want to verify the intuition by rigorous derivations. Theoretically, we may only consider the case of two walkers start at nodes $a$ and $b$, as one can easily generalize to the case of more walkers.  For a walker starts at node $a$, we define $G_{aj}(t)$ as the probability that node $j$ has not been visited by walker $a$ till time $t$. Then the GMFPT is $<\bar{T}>_{ab}=\frac{1}{N}\sum_{j\in V}\sum_{t=0}^\infty G_{aj}(t)G_{bj}(t)$. And the overlap is defined as $O(a,b)=\sum_{j\in V}\sum_{t=0}^{\infty} \left(P_{aj}(t)-\frac{k_j}{\sum k}\right)\left(P_{bj}(t)-\frac{k_j}{\sum k}\right)$. We claim that they are positively correlated with each others. We start from the deep relation between $P_{ij}(t)$ and $G_{ij}(t)$ \cite{PhysRevLett.92.118701}. Applying the discrete Laplace transform, we have, $\widetilde{G}_{ij}(z)=\left(\widetilde{P}_{ij}(z)-\widetilde{P}_{jj}(z)-\delta_{ij}\right)/\left((z-1)\widetilde{P}_{jj}(z)\right)$. Then using the discrete Laplace transform formula of a product, we can write the GMFPT and the overlap in terms of $\widetilde{G}_{ij}(z)$ and $\widetilde{P}_{ij}(z)$ as, $<\bar{T}>_{ab}=\frac{1}{N}\sum_{j\in V}\mathcal{L}\{G_{aj}G_{bj}\}(1)=\frac{1}{N} \sum_{j\in V}\oint_{C}\widetilde{G}_{aj}(\tau)\widetilde{G}_{bj}(\tau^{-1})$ $\tau^{-1}\mathrm{d}\tau$, and, $O(a,b)=\sum_{j\in V}\oint_{C}\left(\widetilde{P}_{aj}(\tau)-\frac{k_j}{\sum k}\frac{1}{1-\tau^{-1}}\right)\left(\widetilde{P}_{bj}(\tau^{-1})-\frac{k_j}{\sum k}\frac{1}{1-\tau}\right)$$\tau^{-1}\mathrm{d}\tau$. Note that the part which only depends on $j$ will become a constant under the summation over $j$ in both equations. We also find that $1/\widetilde{P}_{jj}(z)$ is holomorphic in the region enclosed by $C$ and its value on the positive real axis is always real and positive. By these facts we conclude that $\widetilde{G}_{ij}(z)$ and $\widetilde{P}_{ij}(z)$ are positively correlated with each others. 
	
	Following the analysis above, we may focus on minimizing the total overlap to optimize the GMFPT. This is the starting point of the genetic algorithm introduced in this paper, in which we define the total overlap as the fitness function. Computing the mutual overlap is faster and only requires local topological structures in the surrounding region of initial nodes if we imposed an appropriate cutoff time.
	%------------------------------------------------------------------------------
	\section{Mutation Only Genetic Algorithm}
	\label{sect:algorithm}	
	We start with formally defining the optimization problem as "the optimal set of initial nodes" problem. The input is an undirected network $G$ with size $|V|$ represented by the adjacency list and the number of random walkers $r$. While the output is the optimal initial nodes of these $r$ random walkers. To construct the set of $r$ initial nodes, we may start from randomly choosing a node and iteratively run the genetic algorithm to decide a new initial node which has minimum overlap with the existed ones, and then add it to the set. Following this strategy, the search space of the genetic algorithm is reduced to the set of nodes $V$.
	
	We use a special class of genetic algorithm, the mutation only genetic algorithm (MOGA), (note that it does not stand for the multi-objective genetic algorithm), which defines a more general formulation of mutation. Suppose there are $N$ chromosomes of $L$-bits in the decreasing order of fitness. We assign different mutation probabilities to each locus and they form the mutation matrix $\mathbf{M}$ of size $N\times L$, where $M_{ij}$ is the probability to mutate the $j$-th locus of the $i$-th chromosome.
	
	The design of chromosome representation requires the definition of a one-to-one correspondence between a set of $L$-bit binary chromosomes and $V$. However, if a chromosome is encoded with the degree and the clustering coefficient of the node, since they are not contiguous on the network, we cannot easily find the new node which is represented by the mutated chromosome. An approach to solve this problem is to define the chromosome as a sequence of characters of the nodes on a specific path from a preset node. We thus only need to search among the nearest neighbors to find the new node correspond to the mutated chromosome. And we can obtain the candidate node from a path by picking out the last node. A difficulty of this representation is that the chromosomes may have variable length. However, we can easily extend the length of a chromosome by filling the remaining space with the characters of the last node in the path. In practice, we extend all chromosomes to a fixed length, which is larger than the diameter $d_{\text{max}}$ of the network $G$.
	
	The design of mutation matrix is special since our chromosomes represent paths. To achieve effective exploration and exploitation at the same time, we should only mutate the last several loci of a fit chromosome to perturb the end of the path and search the surrounding area of the current node, while mutate nearly all the loci of an unfit chromosome, so as to generate new paths for search in the region far from the current solution. Since there is no fixed meaning of a specific locus, the mutation probabilities should not depend on the statistics of loci. We define the mutation matrix as,
	\small
	\newcommand\x{1/2}
	\newcommand\bigzero{\makebox(0,0){\text{\huge0}}}
	\newcommand*{\bord}{\multicolumn{1}{c|}{}}
	\[
	\mathbf{M} = \left(
	\begin{array}{ccccc}
	& & & & \bord \\ \cline{5-5}
	& \bigzero & & \bord & \x \\ \cline{4-4}
	& & \bord & \cdots & \x \\ \cline{3-3}
	& \bord & \cdots & \cdots & \cdots \\ \cline{2-2}
	\bord& \x & \x & \cdots & \x \\ \cline{1-1}
	\end{array}\right)
	\]
	\normalsize
	The special triangular form with each element being $1/2$ perturbs the ends of fit paths and uniformly generates new paths simultaneously. Numerical results verify that the special design of the chromosomes and the mutation matrix helps the genetic algorithm provide solutions closed to the optimum in reasonable time.
	%------------------------------------------------------------------------------
	\section{Numerical Results and Conclusions}
	\label{sect:simulations_conclusions}
\begin{figure}[ht]
		\centering
		\subfloat[BA networks with size $|V|=200$ or $800$, and average degree $\left<k\right>=4$ or $10$.]{%
			\includegraphics[width=0.4\textwidth]{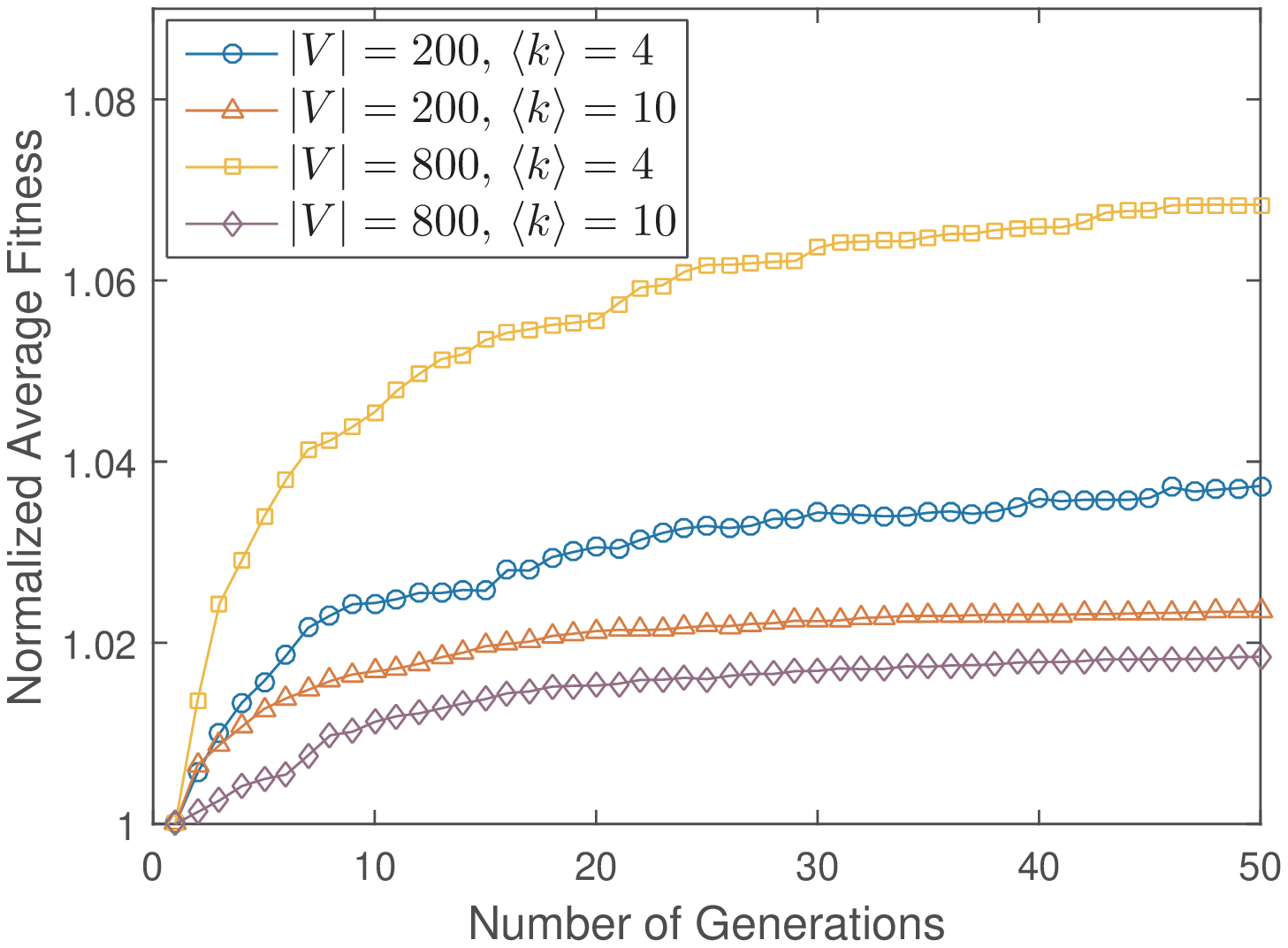}
			\label{fig:fitness_BA}
		}%
		\qquad
		\subfloat[WS networks with size $|V|=500$, average degree $\left<k\right>=6$, and $\beta=0.3$, $0.5$, $0.7$ or $0.9$]{%
			\includegraphics[width=0.4\textwidth]{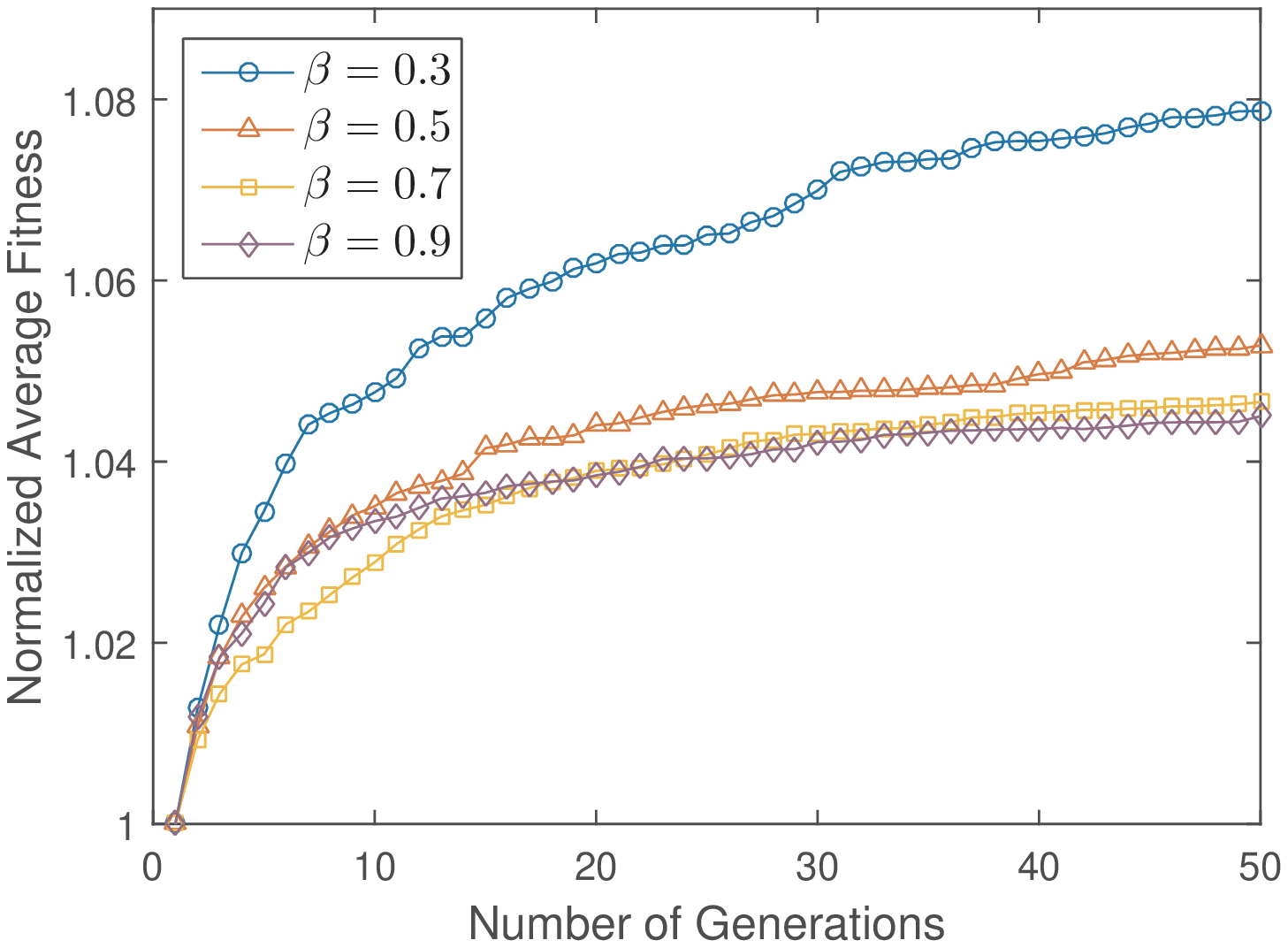}
			\label{fig:fitness_WS}
		}%
		\caption{\small\label{fig:fitness}Normalized average fitness versus the number of generations when running genetic algorithm on BA and WS networks. The fitness is normalized with respected to its initial value and averaged on 30 independent experiments.}
	\end{figure}

	We run the genetic algorithm on two kinds of random networks (Watts-Strogatz and Barab\'asi-Albert). Firstly, we evaluate the performance of our algorithm by computing the fitness of the fittest chromosome in each generation and draw Figure \ref{fig:fitness}. We vary the size $|V|$ and the average degree $\left<k\right>$ of BA networks and the rewiring probability $\beta$ of WS networks. From Figure \ref{fig:fitness_BA}, we can see that the average improvement of fitness within $50$ generations does not explicitly depend on $|V|$, but decreases with increasing $\left<k\right>$. From Figure \ref{fig:fitness_WS}, we find that the improvement strictly decreases with respect to $\beta$. Note that a small $\beta$ in WS model leads to a flat degree distribution and a large average path length $\left<d\right>$. By these observations, we conclude that our algorithm has better performance on networks with small average degree $\left<k\right>$ and large average path length $\left<d\right>$. This may be because the GMFPTs of all possible sets of $r$ initial nodes in networks with large $\left<k\right>$ and small $\left<d\right>$ is sharply distributed so that there is limited room for improvement.

\begin{figure}[ht] 
		\centering
		\subfloat[Path lengths between all pairs of initial nodes]{%
			\includegraphics[width=0.32\textwidth]{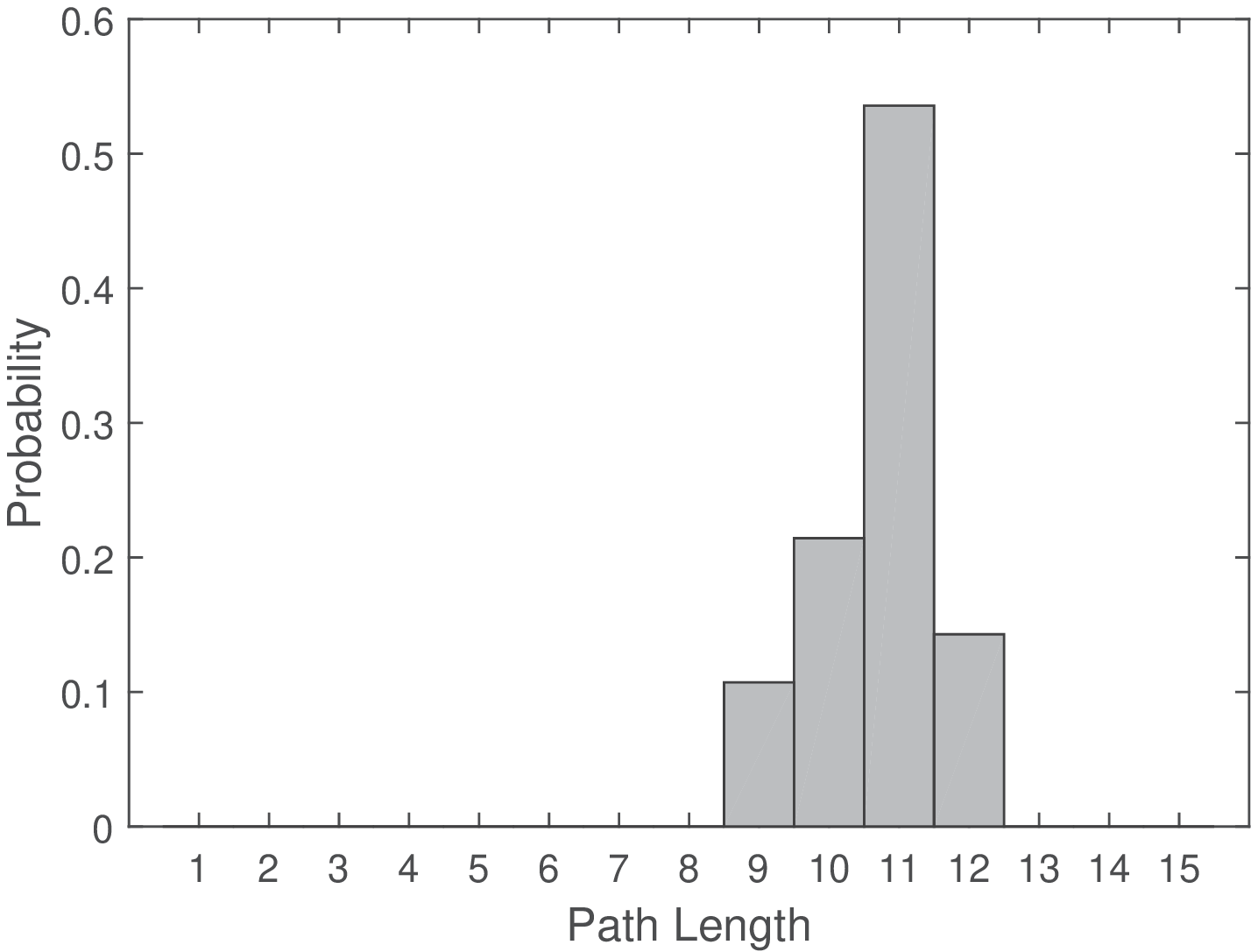}
			\label{fig:path_length_initial_nodes}
		}%
		\qquad\qquad
		\subfloat[Path lengths between all pairs of nodes]{%
			\includegraphics[width=0.32\textwidth]{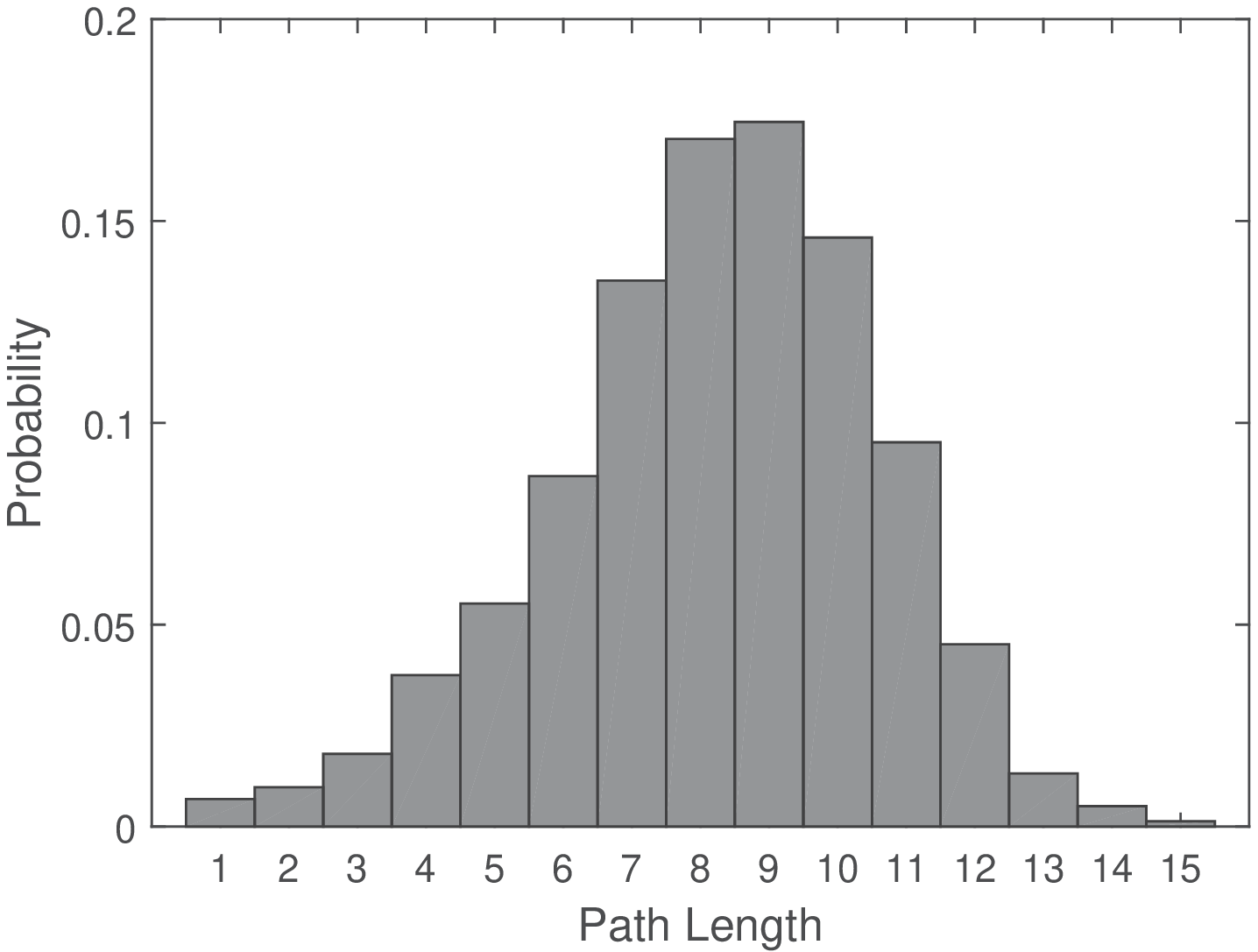}
			\label{fig:path_length_all_nodes}
		}%
		\caption{\small\label{fig:path_length}Normalized path-length histograms of all pairs of nodes in the set of $r=3$ initial nodes produced by the algorithm and in $V$. The algorithm runs on a WS network with size $|V|=600$, average degree $\left<k\right>=4$, and $\beta=0.1$. Distribution in Figure \ref{fig:path_length_initial_nodes} is an average of 20 experiments.}
	\end{figure}

	Secondly, we compute the path lengths between all pairs of initial nodes produced by the algorithm. We expect that every two initial nodes are far from each other so that the total overlap is small. Consider the special case of a regular ring lattice, in which the optimal positions of $r=3$ random walkers should be arranged with equal spacing on the ring so that the path lengths between them are approximately $2d_{\text{max}}/3$, where $d_{\text{max}}$ is the diameter. For WS networks with small $\beta$, which is not far from regular ring lattices, we expect the distribution of path lengths of the set of $r=3$ initial nodes also has a sharp peak at $2d_{\text{max}}/3$. This is verified by numerical results shown in Figure \ref{fig:path_length}. 

	Finally, we calculate the percentage reduction of the GMFPT of the selected set of initial positions compared to the average. Table \ref{tab:improvement} summarizes the reduction of the GMFPT of BA networks with various $\left<k\right>$ and WS networks with various $\beta$.
	\begin{table}[ht]
		\begin{centering}
			\begin{tabular}{ccccccc}
				\hline
				&\multicolumn{3}{c}{BA ($|V|=300$)}&\multicolumn{3}{c}{WS ($|V|=300$,$\left<k\right>=6$)}\\
				GMFPT&$\left<k\right>=4$&$\left<k\right>=6$&$\left<k\right>=8$&$\beta=0.3$&$\beta=0.5$&$\beta=0.7$\\
				\hline
				Improvement&2.6\%&1.5\%&1.0\%&4.9\%&2.2\%&1.8\%\\
				Uncertainty$(\pm)$&0.6\%&0.4\%&0.4\%&0.9\%&0.5\%&0.6\%\\
				\hline
			\end{tabular}
			\caption{\small\label{tab:improvement}The percentage reduction of the GMFPT of $r=3$ initial nodes produced by the algorithm compared to the average. Each result is an average of 20 experiments.}
		\end{centering}
	\end{table}

	We conclude that we have found a method to optimize the global mean first-passage time (GMFPT) by choosing the starting nodes of multiple random walkers to search for a general target in a complex network. As the GMFPT of multiple random walkers depends monotonically on the overlap, we can minimize the total overlap to reduce GMFPT. We achieve this with the mutation only genetic algorithm (MOGA) which finds the initial position of a new walker that has minimum overlap with the existing ones. Special forms of the chromosomes and the mutation matrix are introduced and we achieve balance between exploration and exploitation. Numerical works on WS and BA networks confirm the effectiveness of our method, as shown in Table \ref{tab:improvement}. Our method may be very useful in speeding up searches on large networks. We expect further improvement by means of a more sophisticated MOGA with crossover \cite{law2007adaptive}.
	%------------------------------------------------------------------------------
	% Refs:	
	\label{sect:bib}
	\bibliographystyle{unsrt}
	\bibliography{Reference_Ding_Szeto}
\end{document}